\documentclass[11pt,twoside]{article} 
\usepackage{asp2004}
\usepackage{epsf}\usepackage{psfig}
\usepackage{lscape} 

\begin{document} 
\title{Detailed Spectroscopic and Photometric Analysis of DQ White Dwarfs}
\author{P. Dufour, P. Bergeron, and G. Fontaine} 
\affil{D\'epartement de Physique, Universit\'e de Montr\'eal, C.P. 6128, Succ. Centre-Ville, Montr\'eal, Qu\'ebec, Canada, H3C 3J7}

\begin{abstract} 
We present an analysis of spectroscopic and photometric
data for cool DQ white dwarfs based on improved model
atmosphere calculations. In particular, we revise the atmospheric
parameters of the trigonometric parallax sample of 
\citet{blr01}, and discuss the astrophysical implications on the
temperature scale and mean mass, as well as the chemical evolution of
these stars. We also analyze 40 new DQ stars discovered in the first
data release of the Sloan Digital Sky Survey.
\end{abstract}

\section{Introduction}

Cool white dwarfs showing traces of carbon, either as neutral carbon
lines or molecular C$_2$ Swan bands, are collectively known as DQ
stars. Past analyses showed that these stars are helium rich with
carbon abundances ranging from $\log{\rm C/He}=-7$ to $-2$ as
determined from optical or ultraviolet (IUE) spectroscopic observations
\citep{bues73,grenfell74,koester82,wegner84,weidemann95}.

The presence of carbon has been successfully explained by a model in
which carbon diffusing upward from the core is brought to the
photosphere by the deep helium convection zone
\citep{pelletier86}. This model was shown to reproduce the observed carbon
abundance distribution as a function of effective temperature fairly
well for a helium layer thickness of $\log q({\rm He})\equiv M_{\rm
He}/M_{\star}\sim -3.75$, a value at odds with those obtained from
evolutionary models of post-AGB stars that predict much thicker helium
layers of the order of $\log q({\rm He})\sim -2$.

More recently, \citet{brl97} and \citet[][hereafter collectively
referred to as BLR]{blr01} presented a photometric and spectroscopic
analysis of a large sample of white dwarfs aimed at improving our
understanding of the chemical evolution of cool white dwarfs. Their
sample includes several DQ stars that were analyzed with pure-helium
model atmospheres, the only models available at that time. Recently,
\citet{provencal02} studied the importance of including carbon in the
atmospheres of the DQ star Procyon B. Their effective temperature of
$T_{\rm eff}=7740\pm 50$~K, derived from models including carbon, was
significantly cooler than the earlier value of $T_{\rm eff}=8688\pm
200$~K obtained by
\citet{provencal97} on the basis of pure helium models. This large
difference was attributed to an increase of the He$^-$ free-free opacity
resulting from the additional free electrons provided by carbon.

The natural step following the comprehensive studies of BLR
is thus to include explicitly metals and molecules in the model
atmospheres.

\section{Theoretical Framework}

The model atmosphere code is a modified version of that described at
length in \citet{bergeron95}, which is appropriate for pure hydrogen
and pure helium atmospheric compositions, as well as mixed hydrogen
and helium compositions. In order to treat in detail white dwarfs of
the DQ and DZ spectral types, we had to include metals and
molecules in our equation-of-state and opacity calculations.  Details
will be provided elsewhere (Dufour et al. 2005, in preparation).
Since in this analysis we are only concerned with DQ stars showing
the C$_2$ Swan bands, we need only to include helium and carbon. Our
model grid covers a range of $T_{\rm eff}=5000$ to 12,000~K in steps
of 500~K, $\log g=7.5$ to 9.5 in steps of 0.5 dex, and $\log {\rm
C/He}=-9.0$ to $-2.0$ in steps of 0.5 dex.

\section{Analysis of the BLR Sample}

The first sample used for this study is drawn from the BLR analysis.
It consists of 12 DQ stars with trigonometric parallax measurements,
high signal-to-noise spectroscopy showing the carbon molecular bands,
optical $BVRI$ and infrared $JHK$ photometry. The method used to fit
the data is similar to that described at length in BLR, with the
exception that in addition to the effective temperature and the solid
angle $\pi(R/D)^2$, we now have a third fitting parameter, the carbon
abundance. Briefly, we transform the magnitudes at each bandpass into
average fluxes. The resulting energy distributions are then
fitted with the model fluxes, properly averaged over the filter
bandpasses, using a nonlinear least-squares method.

We begin by assuming a pure helium composition and $\log g= 8.0$ to
get initial estimates of $T_{\rm eff}$ and $\pi(R/D)^2$.  The solid
angle is combined with the distance $D$ derived from the trigonometric
parallax to obtain the stellar radius $R$. This radius is then
converted into $\log g$ using an appropriate mass-radius relation for
white dwarfs, and the procedure is repeated until a convergence in
$\log g$ is reached. Next we turn to the spectroscopic observations
and determine the carbon abundance by fitting the Swan bands at the
values of $T_{\rm eff}$ and $\log g$ obtained from the fit to the
energy distribution. This carbon abundance is then used to obtain new
estimates of the atmospheric parameters from the energy distribution,
and so forth. We repeat the procedure --- typically five iterations
--- until $T_{\rm eff}$, $\log g$, and the carbon abundance
converge to a consistent photometric and spectroscopic solution.  The
$\log g$ values are then converted into masses using the cooling
models of \citet{fon01}.

\begin{figure}[!ht]
\plotone{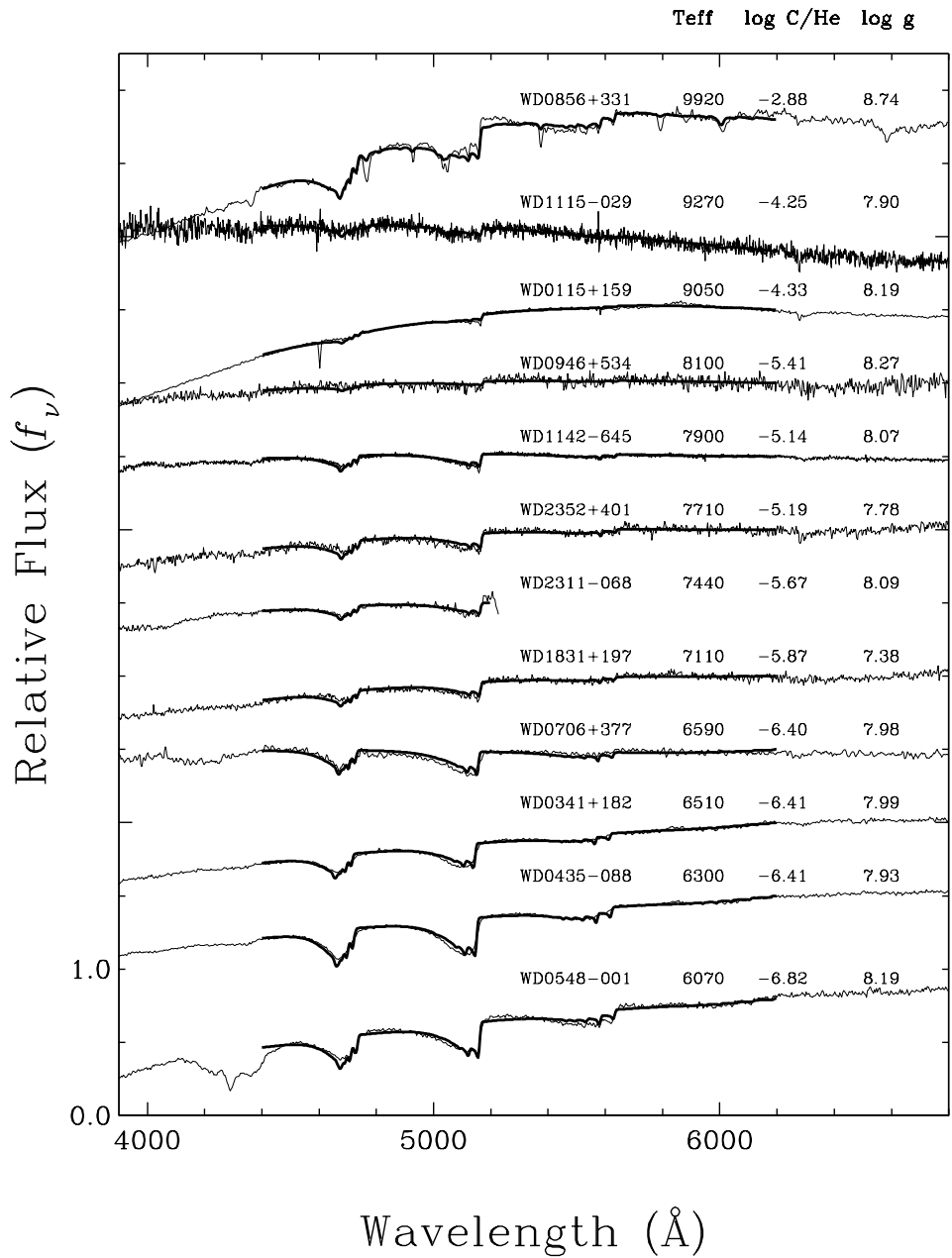}
\caption[FitDQBLR.epsi] {Fits to the Swan bands plotted in order of 
decreasing temperatures from top to bottom. WD~0856+331 also shows
atomic C~\textsc{i} lines.
\label{fg:FitDQBLR}}
\end{figure}

Our spectroscopic fits for the BLR sample are displayed in Figure 
\ref{fg:FitDQBLR}; the photometric fits differ very little from those
already shown in BLR and will thus not be repeated here.

\begin{figure}[!ht]
\plotone{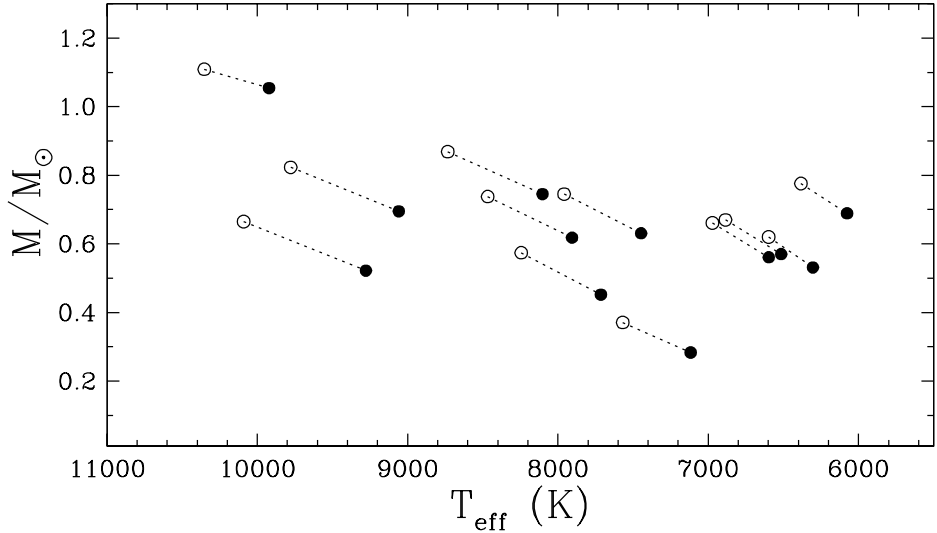}
\caption[shift.epsi] {Comparaison of effective temperatures and masses 
obtained from pure helium models ({\it open circles})
and models including carbon ({\it filled circles}).
\label{fg:shift}}
\end{figure}

We compare in Figure \ref{fg:shift} the effective temperatures and
masses derived from pure helium models (e.g.~BLR) with those obtained
here with models including carbon (see also Weidemann, these
proceedings).  We find that our revised effective temperatures are
reduced significantly with respect to the pure helium solutions, in
agreement with the conclusions of \citet{provencal02} for Procyon
B. The inclusion of carbon in the equation-of-state increases the
number of free electrons and consequently the contribution of the
He$^-$ free-free opacity. The resulting effects on the
atmospheric structures can produce significant differences in the
derived atmospheric parameters. Since the $T_{\rm eff}$ values have
been greatly reduced, so are the model fluxes, and larger solid angles
are required to fit the photometric data, which imply larger stellar
radii or smaller masses, as shown in Figure \ref{fg:shift}.

An interesting consequence of our findings concerns the mass
distribution of DQ stars. An intriguing question raised by BLR was the
fact that the mean mass of DQ stars, $\langle
M\rangle=0.72~M_{\odot}$, was much higher
than that obtained for their likely progenitors, the DB stars, by
\citet{beauchamp96}, $\langle M \rangle=0.59~M_{\odot}$, based on 
spectroscopic fits. Our improved analysis brings the mean mass of DQ
stars down to $\langle M \rangle=0.61~M_{\odot}$,
in much better agreement with the spectroscopic mean mass of DB
stars, and more in line with our understanding of white dwarf evolution.

\section{Analysis of the SDSS Sample}

Our second sample consists of 40 stars spectroscopically identified as
DQ (human ID) in the first data release of the Sloan Digital Sky
Survey (SDSS). Details concerning these observations are described in
\citet{kleinman04} and references therein. Prior to the SDSS study, only 
37 cool ($T_{\rm eff} < 15,000$~K) white dwarfs with carbon features
had been analyzed by various investigators, and the SDSS first data
release {\it alone} has more than doubled the number of DQ stars known,
allowing us to improve significantly the statistics of DQ stars. The
method we use here to fit the SDSS data is similar to that described
above, with the exception that the SDSS $ugriz$ photometry is used
instead.  These photometric passbands cover the entire optical range
from the UV atmospheric cutoff (3200~\AA) to the red sensitivity
cutoff of the detector ($\sim 10,000$~\AA). Furthermore, since
trigonometric parallax measurements are not available for the SDSS
stars, we assume a value of $\log g=8.0$ for all objects.

\section{Discussion}

\begin{figure}[!ht]
\plotone{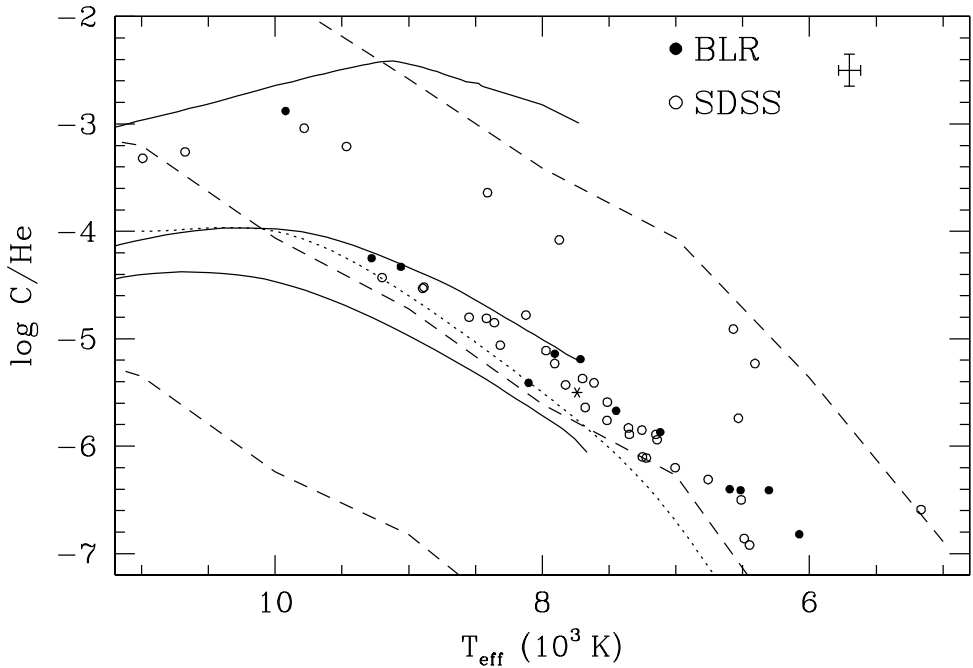}
\caption[pelletier.epsi] {Observed carbon abundances as a function of
$T_{\rm eff}$ for both the BLR sample ({\it filled circles}) and SDSS
sample ({\it open circles}); the star symbol shows the location of
Procyon B. The error bars in the upper right corner indicate the
average uncertainties of $T_{\rm eff}$ ($\sim 150$~ K) and log C/He
($\sim 0.3$ dex). The dotted line represents our detection threshold
for the C$_2$ Swan bands (or atomic C~\textsc{i} lines for $T_{\rm eff}>
10,000$~K). The solid and dashed curves correspond to the evolutionary sequences
discussed in the text.
\label{fg:pelletier}}
\end{figure}

Our results are summarized in Figure \ref{fg:pelletier} where we show
the carbon abundances as a function of $T_{\rm eff}$ for both the BLR
and SDSS samples. We see that the carbon abundances form a relatively
narrow sequence in the $\log {\rm C/He}-T_{\rm eff}$ plane. The
sequence defined by the SDSS sample overlaps the BLR sequence quite
nicely, suggesting that the physical characteristics of these two
samples are very similar. Since the BLR sample as a whole has a normal
mean mass near 0.6 $M_\odot$, the value of $\log g=8$ assumed for the
SDSS sample is probably reasonable. It is also comforting to see that
Procyon B (star symbol at 7740 K), with a precise mass determination
of 0.602 $M_{\odot}$ determined by \citet{provencal02}, falls well
within our sequence. Our results are thus entirely consistent with the
idea that all DQ stars evolved from hotter DB white dwarfs.

Also shown in Figure \ref{fg:pelletier} are the results from two
different sets of evolutionary model calculations, which predict the
carbon abundances as a function of decreasing effective temperatures
for various values of the thickness of the helium layer $q({\rm
He})$. The dashed curves represent the calculations of
\citet{pelletier86} at 0.6 $M_\odot$ with, from top to bottom,
$\log q({\rm He})=-4.0$, $-3.5$, and $-3.0$. As discussed in the
Introduction, our abundance determinations would suggest values of
$\log q({\rm He})$ smaller than $-3.5$ for all DQ stars, much thinner
than the values predicted by post-AGB evolutionary models. A new
generation of evolutionary models computed by Fontaine \& Brassard
(these proceedings) yield quite different results, however, as shown
in Figure \ref{fg:pelletier} by the solid curves for models at 0.6
$M_\odot$ with, from top to bottom, $\log q({\rm He})=-4.0$, $-3.0$,
and $-2.0$. With these new models, the observed carbon abundances in
DQ stars are consistent with values of the helium layer thickness
between $\log q({\rm He})=-3$ and $-2$, in much better agreement with
the results from post-AGB models.

Several objects in Figure \ref{fg:pelletier} show carbon abundances
that lie about 1 dex above the bulk of DQ stars. Among these, the only
white dwarf with a measured trigonometric parallax is G47-18
(WD~0856+331; top filled circle) with an estimated mass of
$M=1.05~M_{\odot}$, by far the most massive DQ star in Figure
\ref{fg:FitDQBLR}. It is thus tantalizing to suggest that the DQ
stars with larger-than-average carbon abundances represent the high
mass component of the white dwarf mass distribution.
\citet{Liebert2003} has suggested that the hot DQ stars discovered
in the SDSS could all be massive, and could correspond to the missing
high mass tail of the DB star mass distribution \citep{beauchamp96}. If
both hypotheses are confirmed, the cool DQ stars with large carbon
abundances observed here would simply represent the natural extension
of the hot DQ stars identified by Liebert et al. Finally, the absence
of neutral carbon lines in the optical spectra of DB stars with normal
masses is naturally explained by our results. Indeed, the expected
carbon abundances in hot ($T_{\rm eff} > 10,000$~K) helium-rich white
dwarfs with $\log q({\rm He})$ between $-3$ and $-2$ fall below the
detection threshold of the atomic C~\textsc{i} lines according to Figure
\ref{fg:pelletier}.

This work was supported in part by the NSERC Canada and by the Fund
FQRNT (Qu\'ebec).


\begin{thebibliography}{}

\bibitem[Beauchamp et al.(1996)]{beauchamp96} Beauchamp, A., 
Wesemael, F., Bergeron, P., Liebert, J., \& Saffer, R.~A.\ 1996, ASP 
Conf.~Ser.~96: Hydrogen Deficient Stars, 295 

\bibitem[Bergeron et al.(2001)]{blr01} Bergeron, 
P., Leggett, S.~K., \& Ruiz, M.~T.\ 2001, \apjs, 133, 413

\bibitem[Bergeron et al.(1997)]{brl97} Bergeron, 
P., Ruiz, M.~T., \& Leggett, S.~K.\ 1997, \apjs, 108, 339

\bibitem[Bergeron et al.(1995)]{bergeron95} 
Bergeron, P., Saumon, D., \& Wesemael, F.\ 1995, \apj, 443, 764 

\bibitem[Bues(1973)]{bues73} Bues, I.\ 1973, \aap, 28, 181 

\bibitem[Fontaine et al.(2001)]{fon01} Fontaine, G., Brassard, P., 
\& Bergeron, P. 2001, \pasp, 113, 409

\bibitem[Grenfell(1974)]{grenfell74} Grenfell, T.~C.\ 1974, \aap, 
31, 303 

\bibitem[Kleinman et al.(2004)]{kleinman04} Kleinman, S.~J., et 
al.\ 2004, \apj, 607, 426 

\bibitem[Koester et al.(1982)]{koester82} 
Koester, D., Weidemann, V., \& Zeidler, E.-M.\ 1982, \aap, 116, 147 

\bibitem[Liebert et al.(2003)]{Liebert2003} Liebert, J., Harris,
H.~C., Schmidt, G.~D., et al.\ 2003, \aj, 126, 2521 

\bibitem[Pelletier et al.(1986)]{pelletier86} Pelletier, C., 
Fontaine, G., Wesemael, F., Michaud, G., \& Wegner, G.\ 1986, \apj, 307, 
242 

\bibitem[Provencal et al.(2002)]{provencal02} Provencal, J.~L., 
Shipman, H.~L., Koester, D., Wesemael, F., \& Bergeron, P.\ 2002, \apj, 
568, 324 

\bibitem[Provencal et al.(1997)]{provencal97} Provencal, J.~L., 
Shipman, H.~L., Wesemael, F., Bergeron, P., Bond, H.~E., Liebert, J., \& 
Sion, E.~M.\ 1997, \apj, 480, 777 

\bibitem[Wegner \& Yackovich(1984)]{wegner84} Wegner, G.~\& 
Yackovich, F.~H.\ 1984, \apj, 284, 257 

\bibitem[Weidemann \& Koester(1995)]{weidemann95} Weidemann, V.~\& 
Koester, D.\ 1995, \aap, 297, 216 

\end{thebibliography}
\end{document}